\documentstyle[12pt]{article}

\def\Sig{\Sigma}
\def\del{\partial}
\def\nab{\nabla}
\def\til{\tilde}
\def\dis{\displaystyle}

\def\a{\rm a}
\def\b{\rm b}

\begin{document}


\begin{center}
{\large\bf Refined form of the paper on the canonical formalism of the 
$f(R)$-type gravity in terms of Lie derivatives}
\\[7mm]
\end{center}

\hspace*{15mm}\begin{minipage}{13.5cm}
Y. Ezawa$^a$ and Y. Ohkuwa$^b$\\[3mm]
$^a$Department of Physics, Ehime University, Matsuyama, 790-8577, Japan\\[1mm]
$^b$Section of Mathematical Science, Department of 
Social Medicine, Faculty of Medicine, University of Miyazaki, Kiyotake, 
Miyazaki, 889-1692, Japan\\[1mm]
{\small Email :  ezawa@sci.ehime-u.ac.jp, 
ohkuwa@med.miyazaki-u.ac.jp} 
\\[5mm]
{\bf Abstract}\\[2mm]
We refine the presentation of the previous paper of our group, Y.Ezawa et al.,
Class. and Quantum Grav. {\bf 23} (2006), 3205. 
In that paper, we proposed a canonical formalism of f(R)-type generalized 
gravity by using the Lie derivatives instead of the time derivatives.
However, the use of the Lie derivatives was not sufficient.
In this note, we make use of the Lie derivatives as far as possible, 
so that no time derivatives are used and the presentation is largely 
improved.
\end{minipage}

\section{Introduction}

 Since the use of the f(R)-type gravity by Caroll et al.\cite{CDTT} 
to explain the discovered accelerated expansion of the universe
\cite{Accel}, the theory has been attracting much attention and its 
various aspects and applications have been investigated\cite{Rev}.
However, its canonical formalism had not been so systematic.
So in \cite{Lie}, our group proposed a formalism by generalizing the 
canonical formalism of Ostrogradski\cite{Ost}.
The generalization is necessary.
As the scalar curvature $R$ depends on the time derivatives of the 
lapse function and shift vector, these variables have to obey the 
field equations, if the Ostrogradski's method is directly applied.
Then, only the solutions to these equations are allowed for these 
variables.
This, however, is in conflict with general covariance since these 
variables specify the coordinate frame so should be taken arbitrarily.
One of the ways to resolve this problem had been given by Buchbinder 
and Lyakhovich(BL method)\cite{BL}.
However, the BL method has an undesirable property that, when the 
generalized coordinates are transformed, Hamiltonian is also 
transformed\cite{Lie}.

So in the previous paper\cite{Lie}, we proposed a canonical formalism 
of the f(R)-type gravity using the Lie derivatives instead of the time 
derivatives, which is a naturaland economical generalization of the 
formalism of Ostrogradski, so remedies the property of the metod by BL 
mentioned above.
However the use of the Lie derivatives was not sufficient, i.e., Lie 
derivatives and time derivatives were used in a mixed way, so some 
expressions are complex.
In this note, we refine the presentation of our previous paper by 
making use of the Lie derivatives as far as possible so that 
expressions are more concise.

\section{Ostrogradski's method}

Before presenting the refined form of the previous paper, we briefly 
describe the method of Ostrogradski\cite{KOS}.
We consider a system with $N$ degrees of freedom, the generalized 
coordinates of which will be denoted as $q^i\ (i=1,2,\cdots,N)$.
The Lagrangian $L$ is assumed to be defined in the $N(n+1)$-dimensional 
gvelocity phase spaceh, the coordinates of which are expressed as 
$$
D^sq^i\ (s=0,1,\cdots,n),\ \ \ {\rm with}\ \ \ D\equiv {d\over dt}, \eqno(1)
$$
and $n$ is the order of the highest time derivative of the generalized 
coordinate $q^i$, so that the Lagrangian $L$ is expressed as 
$L=L(D^sq^i)$. 
It is possible that $n$ is different for different $i$, but we do not 
think of this possibility for simplicity.
Transition to the canonical formalism is given by the Ostrogradski 
transformation(map), which is the straightforward generalization of 
the Legendre transformation(map) for systems described without higher 
time derivatives.
The transformation is given in the following way. 
Consider the variation of the action $\dis S=\int_{t_{1}}^{t_{2}}Ldt$, 
i.e., $\dis \delta S\equiv S[q^i+\delta q^i]-S[q^i]
=\int_{t_{1}}^{t_{2}}\delta L\,dt.\ \ $
Here
$$
\begin{array}{ll}
\delta L&\!\!\!=
\dis \sum_{i=1}^N\sum_{s=0}^n{\del L\over\del D^sq^i}\delta(D^sq^i)
\\[5mm]
\dis &\!\!\!=
\dis \sum_{i=1}^ND\Bigl[\sum_{s=0}^{n-1}\delta(D^sq^i)
\sum_{r=s+1}^n(-1)^{r-s-1}D^{r-s-1}
\Bigl\{{\del L\over\del(D^rq^i)}\Bigr\}\Bigr]
+\sum_{i=1}^N\sum_{s=0}^n(-1)^sD^s\Bigl\{{\del L\over\del(D^sq^i)}
\Bigr\}\delta q^i.
\end{array}                                                        \eqno(2)
$$
The first term in the second line gives the boundary terms in 
$\delta S$.
$n\times N$ velocity variables $D^sq^i(s=0,1,\ldots,n-1)$ are 
transformed (mapped) to the generalized coordinates of the phase 
space and are usually denoted as $q^i_{s}$ (or $Q^i_{s}$) and are 
often called as the new generalized coordinates for $s\geq 1$ in the 
canonical formalism.
The momenta canonically conjugate to them, which we will denote as 
$p^s_{i}$, are defined as the coefficient of the variation of the 
$D^sq^i$, which are transformed to the new generelized coordinate,
 in the boundary terms: that is
$$
p^s_{i}
\equiv \sum_{r=s+1}^n(-1)^{r-s-1}D^{r-s-1}\left\{{\del L\over
\del(D^rq^i)}\right\},\ \ \ (i=1,2,\ldots,N;\ \ s=0,1,\ldots,n-2), \eqno(3)
$$
and the momentum conjugate to $q_{n-1}^i$ is defined as
$$
p^{n-1}_{i}\equiv {\del L\over\del\dot{q}^i_{n-1}}
={\del L\over\del(D^nq^i)}.                                        \eqno(4)
$$
Thus the phase space is $2nN$-dimensional.
However, the relations $\dot{q}^i_{s}=q^i_{s+1}$ give 
$N(n-1)$ constraints when $\dot{q}^i_{s}$ are written using the 
conjugate momenta, the dimension of the subspace spanned by 
independent coordinates is $N(n+1)$ as it should be.
The Hamiltonian is defined similarly to the Legendre transformation:
$$
H\equiv \sum_{j=1}^N\sum_{s=0}^{n-1}p_{j}^s\dot{q}^j_{s}
        -L(q^i_{0},q^i_{1},\cdots,q^i_{n-1};\dot{q}^i_{n-1})       \eqno(5)
$$
where $\dot{q}^i_{n-1}$ should be replaced by new generalized 
coordinates and momentum from (4).
Canonical equations of motion are satisfied if the Euler-Lagrange 
equations are satisfied.
Euler-Lagrange equations are given by setting the coefficients of 
$\delta q^i$ to vanish in the second term of the second line of (2) 
for each $i$.
It is easily seen that (3)$-$(5) reduce to the Legendre transformation 
for $n=1$.
Finally we note that the boundary terms in the variation of the 
action vanish by requiring the vanishing of the variations of new 
generalized coordinates at the boundaries, which can be consistently 
imposed since the Euler-Lagrange equations are $2n$-th order 
differential equations if $L(D^sq^i)$ is non-linear in $D^nq^i$.

\section{Action of $f(R)$-type gravity}

We start from the following action of the generalized 
gravity of $f(R)$-type;
$$
S=S_{G}+S_{M}=\int d^4x\sqrt{-g}f(R)+S_{M},                        \eqno(6)
$$
where $S_{M}$ is the action of matters. 
In other words, the Lagrangian density for gravity ${\cal L}_{G}$ 
is expressed as
$$
{\cal L}_{G}=\sqrt{-g}f(R).                                        \eqno(7)
$$
As the variables for gravity, we adopt the ADM variables\cite{ADM}.
Then, as noted in the introduction, the scalar curvature, $R$, is 
expressed in terms of the Lie derivatives instead of the time 
derivatives as follows:
$$
R
=h^{ij}{\cal L}_{n}^{\;2}h_{ij}
+{1\over4}\left(h^{ij}{\cal L}_{n}h_{ij}\right)^2
-{3\over4}h^{ik}h^{jl}{\cal L}_{n}h_{ij}{\cal L}_{n}h_{kl}
+{}^3\!R-2N^{-1}\Delta N,                                          \eqno(8)
$$
where $h_{ij}$ is the metric of the hypersurface $\Sig_{t}$ which 
has the normal vector field $n^{\mu}=N^{-1}(1,-N^i)$, $N$ is the 
lapse function and $N^i$ is the shift vector. 
${\cal L}_{n}$ represents the Lie derivative along the normal vector 
field $n$.
${}^3R$ is the scalar curvature of $\Sig_{t}$.
From (7) and (8), ${\cal L}_{G}$ depends on the ADM variables in 
the following way:
$$
{\cal L}_{G}
={\cal L}_{G}(N,h_{ij},{\cal L}_{n}h_{ij},{\cal L}_{n}^{\;2}h_{ij}).\eqno(9)
$$

\section{Variation of the action}

From the expression (9), variation of ${\cal L}_{G}$ is expressed as
$$
\delta{\cal L}_{G}
={\delta{\cal L}_{G}\over\delta N}\delta N
+{\delta{\cal L}_{G}\over\delta h_{ij}}\delta h_{ij}
+{\del{\cal L}_{G}\over\del({\cal L}_{n}h_{ij})}\delta{\cal L}_{n}h_{ij}
+{\del{\cal L}_{G}\over\del({\cal L}_{n}^{\;2}h_{ij})}
\delta{\cal L}_{n}^{\;2}h_{ij}.                                    \eqno(10)
$$
Here  of the first two terms on the right-hand 
are not the partial derivatives but the functional derivatives
since the scalar curvature $R$ depends on the derivatives of $N$ 
and $h_{ij}$ in $\Delta N$ and ${}^3R$ as seen in (8).
Actual calculation is made easier when concrete form (8) is used.
Then we have
$$
\delta{\cal L}_{G}
=\delta\sqrt{h}Nf(R)+\sqrt{h}\delta Nf(R)+\sqrt{h}Nf'(R)\delta R,  \eqno(11)
$$
where
$$
\left\{\begin{array}{lcl}
\delta\sqrt{h}&=&\dis {1\over2}\sqrt{h}h^{ij}\delta h_{ij}, \\[3mm]
\delta R
&=&h^{ij}{\cal L}_{n}^2\delta h_{ij}
+(h^{ij}K-3K^{ij}){\cal L}_{n}\delta h_{ij}
+\left[-h^{ik}h^{jl}({\cal L}_{n}+K)({\cal L}_{n}k_{kl})
       +6K^{il}K^{j}_{\ l}\right]\delta h_{ij}
\\[3mm]
&&+\delta\, {}^3\!R+2N^{-2}\Delta N\delta N-2N^{-1}\Delta\delta N,
\end{array}\right.                                                 \eqno(12)
$$
where $K_{ij}$ is the extrinsic curvature defined as
$$
Q_{ij}\equiv {1\over2}{\cal L}_{n}h_{ij}=K_{ij}.                   \eqno(13)
$$
Note that $\delta{\cal L}_{n}h_{ij}={\cal L}_{n}\delta h_{ij},\ 
\delta{\cal L}_{n}^{\;2}h_{ij}={\cal L}_{n}^{\;2}\delta h_{ij}$
and also

\newpage

$$
\left\{\begin{array}{lcl}
\sqrt{h}Nf'(R)\,\delta\,{}^3\!R&=&-\sqrt{h}Nf'(R)R^{ij}\delta h_{ij}
\\[3mm]
&&+\sqrt{h}\left[\left(Nf'(R)\right)^{;i}h^{kl}\Gamma^j_{kl}
-\left(Nf'(R)\right)^{;l}h^{ik}\Gamma^j_{kl}\right]\delta h_{ij}
\\[3mm]
&&+\del_{k}\left[\sqrt{h}\left\{h^{ik}\left(Nf'(R)\right)^{;j}
-h^{ij}\left(Nf'(R)\right)^{;k}\right\}\right]\delta h_{ij}
\\[3mm]
&&+\del_{i}\Bigl[\sqrt{h}Nf'(R)\left(h^{kl}\,\delta\Gamma^{i}_{kl}
             -h^{il}\,\delta\Gamma^{k}_{lk}\right)\Bigr]
\\[3mm]
&&-\del_{i}\Bigl[\sqrt{h}\left\{h^{ij}\left(Nf'(R)\right)^{;k}\delta h_{kj}
-\left(Nf'(R)\right)^{;i}h^{kl}\delta h_{kl}\right\}\Bigr],
\\[3mm]
\sqrt{h}f'(R)\Delta\delta N&=&
\sqrt{h}\Delta f'(R)\delta N
+\del_{k}\left[\sqrt{h}(f'(R)\nab^k\delta N
              -\nab^kf'(R)\,\delta N)\right] .                     
\end{array}\right.                                                 \eqno(14)
$$
When we use (12) in (11) and apply the variational principle, 
gpartial integrationsh have to be done for terms including 
${\cal L}_{n}\delta h_{ij}$ and ${\cal L}_{n}^{\;2}\delta h_{ij}$.
This is done by using a relation for a scalar field $\Phi$:
$$
{\cal L}_{n}(\sqrt{h}N\Phi)
={\cal L}_{n}(\sqrt{h}N)\,\Phi+\sqrt{h}\,N{\cal L}_{n}\Phi,\ \ \ 
{\cal L}_{n}\Phi=n^{\mu}\del_{\mu}\Phi.                            \eqno(15)
$$
Then we have
$$
\begin{array}{ll}
\sqrt{h}Nf'(R)(h^{ij}K-3K^{ij}){\cal L}_{n}\delta h_{ij}&=
-\sqrt{h}N({\cal L}_{n}+K)\left[f'(R)(h^{ij}K-3K^{ij})\right]\delta h_{ij}
\\[3mm]
&\ \ \ \ +\del_{\mu}\left[n^{\mu}\sqrt{h}Nf'(R)
                          (h^{ij}K-3K^{ij})\delta h_{ij}\right],
\end{array}                                                        \eqno(16)
$$
and
$$
\begin{array}{ll}
\sqrt{h}Nf'(R)h^{ij}{\cal L}_{n}^{\;2}\delta h_{ij}&=
\sqrt{h}N\left[K^2f'(R)h^{ij}+{\cal L}_{n}^{\;2}(f'(r)h^{ij})
+K{\cal L}_{n}(f'(R)h^{ij})\right]\delta h_{ij}
\\[3mm]
&\ \ \ \ +\del_{\mu}\left[n^{\mu}\sqrt{h}N\{f'(R)h^{ij}
\delta{\cal L}_{n}h_{ij}
-({\cal L}_{n}+K)(f'(R)h^{ij})\delta h_{ij}\}\right].                
\end{array}                                                        \eqno(17)
$$
Using these relations, we have for $\delta{\cal L}_{G}$ the 
following expression:
$$
\begin{array}{ll}
\delta{\cal L}_{G}&
=\sqrt{h}\left[\Delta f'(R)+2N^{-1}\Delta Nf'(R)\right]\delta N
+\del_{i}\left[\sqrt{h}\left(f'(R)\nab^i\delta N
-\nab^if'(R)\delta N\right)\right]
\\[3mm]
&+\Bigl[f'''(R)({\cal L}_{n}R)^2h^{ij}
+f''(R)\left({\cal L}_{n}^{\;2}R\,h^{ij}-{\cal L}_{n}K^{ij}\right)
\\[3mm]
&\dis+f'(R)\Bigl(KK^{ij}-h^{ij}{\cal L}_{n}K-h^{ik}h^{jl}{\cal L}_{n}K_{kl}
+6K^{ik}K^j_{\ k}
+{\delta\,{}^3\!R\over \delta h_{ij}}\Bigr)
+{1\over2}f(R)h^{ij}\Bigr]\delta h_{ij}
\\[5mm]
&+\del_{\mu}\left[n^{\mu}\sqrt{h}N\left\{
f'(R)h^{ij}\delta{\cal L}_{n}h_{ij}
-\left(f''(R){\cal L}_{n}Rh^{ij}+f'(R)K^{ij}\right)\delta h_{ij}
\right\}\right].
\end{array}                                                        \eqno(18)
$$

\section{New generalized coordinates and momenta canonically conjugate 
to them}

New generalized coordinates, denoted as $Q_{ij}$, are taken, as 
in \cite{BL}, to be (a half of) the Lie derivatives of the original 
generalizewd coordinates $h_{ij}$ which is equal to the extrinsic 
curvature (13).
Momenta canonically conjugate to the original and new generalized 
coordinates, $p^{ij}$ and $P^{ij}$ respectively, are defined to be 
the coefficient of their variations in the total time derivative 
terms in (17):
$$
\left\{\begin{array}{l}
p^{ij}=-\sqrt{h}\left[{\cal L}_{n}f'(R)h^{ij}+f'(R)Q^{ij}\right],
\\[3mm]
P^{ij}=2\sqrt{h}f'(R)h^{ij},
\end{array}\right.                                                 \eqno(19)
$$
where, of course, ${\cal L}_{n}f'(R)$ is also expressed as 
$f''(R){\cal L}_{n}R$.
Expressions of these equations that correspond to (10) read as follows
\footnote{If we use a scalar function $\til{\cal L}_{G}$ defined as 
${\cal L}_{G}\equiv N\sqrt{h}\til{\cal L}_{G}$, factors $n^0$ disappear 
and we have
$$
p^{ij}=\sqrt{h}{\del\til{\cal L}_{G}\over\del({\cal L}_{n}h_{ij})}
-{\cal L}_{n}\Bigl(\sqrt{h}{\del\til{\cal L}_{G}\over
\del({\cal L}_{n}^{\;2}h_{ij})}\Bigr),\ \ \ 
P^{ij}
=\sqrt{h}{\del\til{\cal L}_{G}\over\del({\cal L}_{n}^{\;2}h_{ij})}.
$$
}:
$$
p^{ij}
=n^0{\del{\cal L}_{G}\over \del({\cal L}_{n}h_{ij})}
-{\cal L}_{n}\Bigl(n^0{\del{\cal L}_{G}\over
\del({\cal L}_{n}^{\;2}h_{ij})}\Bigr),\ \ \ 
P^{ij}=2n^0{\del{\cal L}_{G}\over \del({\cal L}_{n}^{\;2}h_{ij})}. \eqno(20\a)
$$
Or reversingly, we have
$$
{\del{\cal L}_{G}\over \del({\cal L}_{n}h_{ij})}
={1\over n^0}\Bigl(p^{ij}+{1\over2}{\cal L}_{n}P^{ij}\Bigr),\ \ \ 
{\del{\cal L}_{G}\over \del({\cal L}_{n}^{\;2}h_{ij})}
={1\over 2n^0}P^{ij}.                                              \eqno(20\b)
$$

\section{Hamiltonian density}

Correspondence of each point on different $\Sig_{t}$ are given by 
a 1-parameter transformation along the timelike curve for which the 
vector field $t^{\mu}$ is the tangent, so we have, e.g., 
$$
h_{ij}({\bf x},t+\delta t)
=h_{ij}({\bf x},t)+{\cal L}_{t}h_{ij}\delta t.                     \eqno(21)
$$
Actually, we have ${\cal L}_{t}h_{ij}=\del_{0}h_{ij}$ in the coordinate 
frame we are using.
Thus Hamiltonian density ${\cal H}_{G}$ is defined to be
$$
{\cal H}_{G}\equiv 
p^{ij}{\cal L}_{t}h_{ij}+P^{ij}{\cal L}_{t}Q_{ij}-{\cal L}_{G}.    \eqno(22)
$$
${\cal H}_{G}$ has the following form;
$$
{\cal H}_{G}=N{\cal H}_{0}+N^i{\cal H}_{i}+{\rm divergent\ term},  \eqno(23)
$$
where, after a canonical transformation $(Q,P)\rightarrow 
(\bar{Q},\bar{P})\equiv (P,-Q)$, we have
$$
\left\{\begin{array}{lcl}
{\cal H}_{0}
&=&\dis {2\over Q}p^{\dagger ij}p^{\dagger}_{ij}-{2\over d}Pp
+{1\over2}Q\psi(Q/2\sqrt{h})-{d-3\over 2d}QP^2-{1\over2}\,{}^3\!RQ
\\[5mm]
&&-\sqrt{h}f\left(\psi(Q/2\sqrt{h})\right)+\Delta Q
\\[5mm]
{\cal H}_{k}&=&\dis 2p^{\dagger\ \ ;j}_{ij}-{2\over d}p_{;i}
+{2\over d}(QP)_{i}
\end{array}\right.                                                 \eqno(24)
$$

\section{Invariance of the Hamiltonian}

We consider the following transformations of the generalized 
coordinates $h_{ij}$:
$$
h_{ij}\rightarrow \phi_{ij}\equiv F_{ij}(h_{kl})\ \ \ 
{\rm or\ inversely}\ \ \ h_{ij}\equiv G_{ij}(\phi_{kl}),           \eqno(25)
$$
and show that the Hamiltonian is invariant under this transformation.
New generalized coordinates $\Phi_{ij}$ are defined as in (13), i.e.,
$$
\Phi_{ij}\equiv {1\over2}{\cal L}_{n}\phi_{ij}.                    \eqno(26)
$$
Hamiltonian density $\bar{\cal H}_{G}$ expressed in the transformed 
variables is defined to be
$$
\bar{\cal H}_{G}\equiv 
\pi^{ij}{\cal L}_{t}\phi_{ij}+\Pi^{ij}{\cal L}_{t}\Phi_{ij}
-\bar{\cal L}_{G}
(N,\phi_{ij},{\cal L}_{n}\phi_{ij},{\cal L}_{n}^{\;2}\phi_{ij}),   \eqno(27)
$$
where $\pi^{ij}$ and $\Pi^{ij}$ are momenta canonically conjugate 
to $\phi_{ij}$ and $\Phi_{ij}$, respectively, and since
$$
{\cal L}_{n}h_{ij}
={\del G_{ij}\over \del\phi_{kl}}{\cal L}_{n}\phi_{kl},\ \ \ 
{\cal L}_{n}^{\;2}h_{ij}
={\cal L}_{n}\Bigl({\del G_{ij}\over \del\phi_{kl}}\Bigr)
{\cal L}_{n}\phi_{kl}+{\del G_{ij}\over \del\phi_{kl}}
{\cal L}_{n}^{\;2}\phi_{kl},                                       \eqno(28)
$$
$\bar{{\cal L}_{G}}$ is defined as
$$
\bar{\cal L}_{G}(N,\phi_{ij},{\cal L}_{n}\phi_{ij},{\cal L}_{n}^{\;2}h_{ij})
\equiv {\cal L}_{G}\left(N,G_{ij}(\phi_{kl}),{\del G_{ij}\over 
\del\phi_{kl}}{\cal L}_{n}\phi_{kl},{\cal L}_{n}\Bigl({\del G_{ij}\over 
\del\phi_{kl}}\Bigr){\cal L}_{n}\phi_{kl}+{\del G_{ij}\over 
\del\phi_{kl}}{\cal L}_{n}^{\;2}\phi_{kl}\right).                    \eqno(29)
$$
$\pi^{ij}$ and $\Pi^{ij}$ satisfy relations similar to (20a,b), 
and from these relations, we have
$$
\pi^{ij}=p^{kl}{\del G_{kl}\over \del\phi_{ij}},\ \ \ 
\Pi^{ij}=P^{kl}{\del G_{kl}\over \del\phi_{ij}},                  \eqno(30\a)
$$
or inversely
$$
p^{ij}=\pi^{kl}{\del F_{kl}\over \del h_{ij}},\ \ \ 
P^{ij}=\Pi^{kl}{\del F_{kl}\over \del h_{ij}}.                    \eqno(30\b)
$$
With help of (30a,b), we have
$$
p^{ij}{\cal L}_{t}h_{ij}
=\pi^{kl}{\del F_{kl}\over \del h_{ij}}{\del F_{ij}\over 
\del\phi_{mn}}{\cal L}_{t}\phi_{mn}=\pi^{ij}{\cal L}_{t}\phi_{ij}. \eqno(31)
$$
Similar relation holds between $P^{ij}$ and $\Pi^{ij}$, so we have
$$
{\cal H}_{G}=\bar{\cal H}_{G}.                                     \eqno(32)
$$
It is noted that the transformation (25) includes the coordinate 
transformation on $\Sig_{t}$.

\section{Summary}

We presented a canonical formalism of $f(R)$-type gravity in terms 
of the Lie derivatives by refining our previous paper\cite{Lie}.
The formalism is a natural and economical generalization of the 
Ostrogradski's formalism. 
Generalization is necessary to assure the invariance of the theory 
under the general coordinate transformation.

\end{document}